\documentclass[aps,prb,twocolumn,superscriptaddress,longbibliography]{revtex4-2}
\usepackage{amsmath,amssymb}
\usepackage[pdftex]{hyperref,graphicx}
\hypersetup{colorlinks = true, urlcolor = blue, linkcolor = blue, citecolor = blue}
\usepackage{physics}
\usepackage{xcolor}
\usepackage{bm}

\begin{document}
\title{Majorana nanowires, Kitaev chains, and spin models}
\author{Haining Pan}
\affiliation{Condensed Matter Theory Center and Joint Quantum Institute, Department of Physics, University of Maryland, College Park, Maryland 20742, USA}
\author{Sankar Das Sarma}
\affiliation{Condensed Matter Theory Center and Joint Quantum Institute, Department of Physics, University of Maryland, College Park, Maryland 20742, USA}

\begin{abstract}
    Motivated by the fact that the idealized Kitaev chain toy model and the experimental semiconductor-superconductor Majorana nanowire can both host the Majorana zero modes, we theoretically investigate the question to what extent the two models are equivalent or similar, using the perspective of the corresponding dual spin models for both.
    We start with the duality between the Kitaev chain and the transverse-field XY spin model through the Jordan-Wigner transformation with the goal of establishing the connection between the Kitaev chain and the nanowire. 
    By applying the Jordan-Wigner transformation to the nanowire, we find that the corresponding bosonic spin model is a generalized spin cluster model, containing 3- and 4-spin terms, with staggered couplings.
    By projecting out the upper band of the bare semiconductor with higher energy, we obtain an effective low-energy spinless system from the spinful nanowire system deep in the topological regime.     
    Finally, we establish the connection between the Kitaev chain and Majorana nanowire by showing that the spinless Kitaev chain can be viewed as the first-order approximation of the spinful Majorana nanowire deep in the topological regime.
\end{abstract}

\maketitle

\section{Introduction}

The Majorana zero modes (MZMs) are non-Abelian Ising anyons with potential application to fault-tolerant topological quantum computing because of the robustness of the quantum degenerate ground state against local perturbations protected by a topological gap~\cite{nayak2008nonabelian,sarma2015majorana}. One prototypical model that was shown to host boundary MZMs is the Kitaev chain~\cite{kitaev2001unpaired}, which contains the nearest-neighbor hopping and $p$-wave superconductivity in a spinless system~\cite{read2000paired}. Given appropriate parameters and conditions, a pair of Majorana zero modes can emerge at both ends of the Kitaev chain by construction. Although, the Kitaev chain is an idealized toy model to show a proof of principle, because spinlessness and $p$-wave superconductivity are hard to find in nature, the model has been influential in focusing attention on non-Abelian MZMs, topological superconductivity, topological quantum phase transition (TQPT), and topological quantum computing using Majorana zero modes. Therefore, realistic experimentally feasible models have been proposed to search for MZMs. The most prevailing scheme used in experiments---see, e.g., Ref.~\onlinecite{aghaee2022inasal} and references therein for a recent effort to experimentally realize Majorana zero modes in semiconductor-superconductor hybrid structures--- is the one-dimensional (1D) semiconductor-superconductor (SM-SC) nanowire~\cite{lutchyn2010majorana,oreg2010helical,sau2010generic,sau2010nonabelian}, where the semiconductor nanowire with a large Rashba-type spin-orbit coupling (SOC) acquires $p$-wave superconductivity from an $s$-wave superconductor due to the proximity effect, and a pair of MZMs should appear at the ends of the nanowire given a sufficiently large magnetic field. This realistic SM-SC system is often referred to as a Majorana nanowire since the MZMs are predicted to arise at the two ends of the nanowire under suitable conditions. However, although both Kitaev chain and Majorana nanowire systems can host MZMs, they are two completely different models which, at first sight, appear to have little to do with each other except for their one-dimensionality and the possible existence of MZMs.  For example, the Kitaev chain is a lattice model made of spinless fermions whereas the nanowire is a continuum system with real spinful electrons.  Also, in the Kitaev chain, $p$-wave superconductivity is assumed \textit{a priori} in the model whereas in the Majorana nanowire the effective topological $p$-wave superconductivity is emergent under the subtle combined effects of induced $s$-wave SC, Rashba SO-coupling, and Zeeman spin splitting.  Historically, the Majorana nanowire proposal arose not motivated by the Kitaev chain idea, but from various 2D proposals involving SO-coupling, $s$-wave SC, and spin splitting~\cite{fu2008superconducting,zhang2008ip,sau2010generic}. After the fact, however, it seems obvious that the two models, Kitaev chain and Majorana nanowire, should be related since both depend on having an effectively spinless $p$-wave SC leading to MZMs localized at the ends of a 1D system. 
However, the extent to which they are generically equivalent to each other remains unclear despite some isolated efforts~\cite{sau2012realizing,cole2017ising,pan2021disorder,dvir2022realization} to connect them through specific properties. 
{In particular, References~\onlinecite{alicea2012new,alicea2011nonabelian} were among the first to have preliminary discussions on possible connections between the Kitaev chain and the Majorana nanowire models.}

{This lack of clarity sometimes leaves a misconception that these two models, Kitaev chain and Majorana nanowire, are always equivalent, which is incorrect, as we will show later under which circumstances these two models are equivalent and when they cannot be mapped into each other. This lack of complete equivalence between the two models is rather obvious if one considers that the Kitaev model is a spinless one band model whereas the nanowire model is a spinful multiband model.  Thus, a claim that the nanowire model can be obtained by taking the continuum limit of the Kitaev chain and projecting onto the low-energy degrees of freedom is technically incorrect since such a low-energy projection is meaningless for the original Kitaev chain model.}
Thus, we are motivated to study the underlying general connection between the Kitaev chain and the SM-SC nanowire try to see whether we can find an exact Majorana mapping between them.
{Mapping both onto effective spin models by eliminating the fermionic degrees of freedom enables a direct one-to-one comparison between the models to figure out how they are connected and how they are not.  This is what we accomplish in the this paper.}

We first recall the duality~\cite{kitaev2001unpaired,kitaev2009topological,greiter20141d} between the Kitaev chain and transverse-field XY model (which can be further tuned to the transverse-field Ising model given appropriate parameters) as this is the origin of the term Ising anyon for MZMs (which is often used~\cite{nayak2008nonabelian}). This mapping of the Kitaev chain to the corresponding transverse-field XY/Ising model is achieved by a nonlocal Jordan-Wigner transformation converting the fermions to bosonic spin operators. For the same reason, to see the connection between the Kitaev chain and SM-SC Majorana nanowire, we should ask first how to map an SM-SC Majorana nanowire to a corresponding spin model through the Jordan-Wigner transformation. However, we find that the resulting spin model is, unlike the relatively simple form of the transverse-field XY model (as for the Kitaev chain), rather complicated --- it is a generalized cluster model containing up to 4-spin interactions with staggered couplings, which do not exist in the mapping of the Kitaev chain to the transverse-field XY model. Thus the connection between the two models appears much more obscure in the spin language than in the fermionic language, where at least, both models at their cores are free fermion ``band'' problems.

We, therefore, approach the SM-SC Majorana nanowire from another angle: The first step is to convert the spinful nanowire into an effective low-energy model by projecting out the upper band of the bare semiconductor with higher energy in momentum space in the limit of the large magnetic field. 
We can then manually construct a spinless model by ignoring the band index and incorporating the superconducting term. 
We find that the form of the effective low-energy Hamiltonian in momentum space is already a spinless $p$-wave superconductor that contains a normal SM part plus an antisymmetric SC part, which is superficially similar to the Kitaev chain. To make a more direct comparison, we expand the Hamiltonian in momentum space in the limit of a large magnetic field, and recover the real space Hamiltonian up to the third order of approximation. We find that the first order of this high-field expansion can be exactly mapped into the Kitaev chain, and the higher order approximation will result in longer-range hoppings and pairings. Therefore, the Kitaev chain can be viewed at least as a first-order approximation of the SM-SC nanowire model at a high magnetic field deep in the topological regime. This also makes physical sense because, in the high-field limit, the low-energy Majorana nanowire projection essentially produces a spinless $p$-wave SC as assumed by construction in the Kitaev model.  The two models obviously cannot be equivalent at a low magnetic field deep in the trivial phase since the nanowire-induced proximity SC is essentially $s$-wave in the trivial phase whereas by construction the SC is $p$-wave in the Kitaev chain.

This paper is organized as follows. In Sec.~\ref{sec:JW}, we state the conventions for the Jordan-Wigner transformation adopted throughout this paper. In Sec.~\ref{sec:KC}, we recapitulate and expand on the duality between the Kitaev chain and the transverse-field XY model through the Jordan-Wigner transformation. Then we generalize the Jordan-Wigner transformation for the spinful model, and directly apply it to the SM-SC Majorana nanowire Hamiltonian to obtain a generalized spin cluster model with staggered couplings in Sec.~\ref{sec:cluster}. Finally, we present the direct mapping between the Kitaev chain and the SM-SC nanowire by projecting out the higher energy band in Sec.~\ref{sec:eff}. Our conclusion is in Sec.~\ref{sec:conclusion}.

\section{Jordan-Wigner transformation}\label{sec:JW}
The Jordan-Wigner transformation~\cite{jordan1928ueber} is a nonlocal transformation that connects a fermionic operator and a series of nonlocal bosonic operators (in $\frac{1}{2}$-spin systems) while keeping the canonical commutation relation the same.  We define the Jordan-Wigner transformation following the conventions as
\begin{equation}\label{eq:JW}
    \begin{split}
        f_i^\dagger&=\prod_{j=1}^{i-1}\left( -\sigma_j^z \right)\sigma_i^+,\\
        f_{i} &=\prod_{j=1}^{i-1}\left( -\sigma_j^z \right)\sigma_i^-,
    \end{split}
\end{equation}
where $f_i$ ($f_{i}^\dagger $) is the fermionic annihilation (creation) operator acting on site $i$ in real space, and $ \sigma_{i}^{+} $ ($ \sigma_{i}^{-} $) is the bosonic ladder operator that raises (lowers) the spin by one unit at site $i$ in $\frac{1}{2}$-spin systems, which can be expressed in terms of the Pauli matrix:
\begin{equation}
    \sigma_{i}^{\pm} =\frac{1}{2}\left(  \sigma_{i}^{x} \pm i  \sigma_{i}^{y}  \right) .
\end{equation}

It is straightforward to verify that the anti commutating relation of the fermionic operators defined in Eq.~\eqref{eq:JW} is also restored by the commuting bosonic spin operators.

For a general noninteracting fermionic Hamiltonian, the bilinear form of a local product of two fermionic operators can be transformed into the product of a series of local spin operators through the Jordan-Wigner transformation, e.g.,
\begin{equation}\label{eq:JW_2}
    \begin{split}
        f_{i}^\dagger f_{i} &=\frac{1}{2} \left( 1+ \sigma_{i}^{z}  \right),\\
        f_{i+1}^\dagger f_{i} &=  \sigma_{i+1}^{+}  \sigma_{i}^{-}  ,\\
        f_{i+1}^\dagger f_{i}^\dagger &= - \sigma_{i+1}^{+} \sigma_{i}^{+},
    \end{split}
\end{equation}
which preserves the locality.

\section{Kitaev chain}\label{sec:KC}
We first briefly review, for the sake of clarity and completeness, the duality between the Kitaev chain and the $s=\frac{1}{2}$ 1D transverse-field XY model~\cite{suzuki1971relationship}, which can be mapped into each other through the Jordan-Wigner transformation of Eq.~\eqref{eq:JW}. The Kitaev chain~\cite{kitaev2001unpaired} is a spinless fermionic lattice tight-binding model with the nearest-neighbor hopping $t$, and $p$-wave superconducting pairing energy $\Delta$, where its Hamiltonian in real space is 
\begin{equation}\label{eq:KC}
    \hat{H}_{\text{KC}}^{(f)}=\sum_{i=1}^{N} -\mu f_{i}^\dagger f_{i} + \left( -t f_{i+1}^\dagger f_{i} +\Delta f_{i+1}^\dagger f_{i}^\dagger +\text{h.c.} \right)
\end{equation}
Here, $\mu$ is the onsite chemical potential, $N$ is the total number of sites in the Kitaev chain, and we take the periodic boundary condition for simplicity.

By defining the Fourier transformation $\tilde{f}_{k}^\dagger = \frac{1}{\sqrt{N}}\sum_j f_{j}^\dagger e^{ikj}$, we can rewrite the Hamiltonian in momentum space as
\begin{equation}\label{eq:KC_k}
    \hat{H}_{\text{KC}}^{(k)}=\sum_k \left( -2t\cos k-\mu \right) \tilde{f}_{k}^\dagger \tilde{f}_{k} - \left( i\Delta \sin k \tilde{f}_{k}^\dagger f_{-k}^\dagger +\text{h.c.} \right).
\end{equation}
To obtain the band structure, we construct the Bogoliubov-de Gennes Hamiltonian in momentum space, $\hat{H}_{\text{KC}}^{(k)}= \frac{1}{2}\sum_{k}\tilde{\bm{f}}_{k}^\dagger  H_{\text{KC}}^{\text{BdG}}(k) \tilde{\bm{f}}_{k} - 2t\cos k-\mu $, where $\tilde{\bm{f}}_{k} =\left( \tilde{f}_{k} , \tilde{f}_{-k}^\dagger  \right)^\intercal$ and  
\begin{equation}
    H_{\text{KC}}^{\text{BdG}}(k)=(-2t\cos k-\mu)\tau_z+2\Delta\sin k \tau_y.
\end{equation}
Therefore, by squaring the BdG Hamiltonian, the band structure for the Kitaev chain in Eq.~\eqref{eq:KC} is $E_{\text{KC}}^2(k)=\left( 2t\cos k +\mu \right)^2 +4\Delta^2\sin^2k$ .

By applying the Jordan-Wigner transformation Eq.~\eqref{eq:JW}, we obtain the corresponding bosonic spin model by substituting Eq.~\eqref{eq:JW_2} into Eq.~\eqref{eq:KC},
\begin{equation}\label{eq:XY}
    \begin{split}
        \hat{H}_{\text{KC}}^{(\sigma)} &= \sum_{i=1}^{N} -\frac{\mu}{2}  \left( 1+ \sigma_{i}^{z}  \right) - \left( t \sigma_{i+1}^{+}  \sigma_{i}^{-} +\Delta\sigma_{i+1}^{+} \sigma_{i}^{+} +\text{h.c.} \right)\\
        &= -\frac{\mu N}{2}- \sum_{i=1}^{N}   \frac{t+\Delta}{2}  \sigma_{i+1}^{x}  \sigma_{i}^{x}   + \frac{t-\Delta}{2}    \sigma_{i+1}^{y}  \sigma_{i}^{y} + \frac{\mu}{2}    \sigma_{i}^{z} ,
    \end{split}
\end{equation} 
which is the $s=\frac{1}{2}$ 1D transverse-field XY model. Note that Eq.~\eqref{eq:XY} may be easily converted to the simpler transverse-field Ising model by making special choices for $t$, $\Delta$, etc., but the generic mapping of the Kitaev chain is always to the transverse-field XY model, a distinction (i.e., Ising or XY) often overlooked in discussions of the Kitaev chain. 

The direct mapping from the Kitaev chain in Eq.~\eqref{eq:KC} to the transverse-field XY model in Eq.~\eqref{eq:XY} indicates that they have the same band structures and wave functions. Furthermore, the two phases in the Kitaev chain also correspond to the two different phases in the transverse-field XY model, where the phase transition happens at the same point. In both cases, there are two gapped phases, separated by a critical point. For the Kitaev chain, the band gap closes at $2\abs{t}=\abs{\mu}$ where two different phases, trivial and topological, exist for $2\abs{t}<\abs{\mu}$ and $2\abs{t}>\abs{\mu}$, respectively. For the transverse-field XY model, if we can reduce it to a transverse-field Ising model~\cite{pfeuty1970onedimensional} by setting $t=\pm\Delta$ (as we can tell that the change of $\Delta$ cannot close the band gap from the fermionic Hamiltonian), the two phases, the paramagnetic and ferromagnetic state, exist for $2\abs{t}<\abs{\mu}$ and $2\abs{t}>\abs{\mu}$, respectively. This paramagnetic to ferromagnetic transition is, however, not a topological quantum phase transition (TQPT)-- it is a standard breaking of the SU(2) spin symmetry at a phase transition. Ising model does not possess Ising anyons!  Despite their dual nature revealed by the Jordan- Wigner transformation converting one to the other, the physics of the two models is qualitatively different. The Kitaev chain manifests a topological phase transition protected by the particle-hole symmetry throughout the entire change of parameters, while the transverse-field XY model is a conventional spontaneous symmetry breaking phase transition, where the spin $\mathbb{Z}_2$ symmetry breaks as $\abs{t}$ increases. The spin symmetry is not protected as any random local magnetic field fluctuation can break it as was already clearly stated in a footnote in Ref.~\onlinecite{kitaev2001unpaired}.

\section{Semiconductor-superconductor nanowire}\label{sec:NW}
All of the above are standard procedures to show the duality between the Kitaev chain and transverse-field XY (or Ising) model via the Jordan-Wigner transformation. 
However, the Kitaev chain is a mere toy model to establish the proof of principle that $p$-wave spinless superconductor can support Majorana zero modes, which was in fact already pointed out in Ref.~\onlinecite{read2000paired} in the context of 2D systems and fractional quantum Hall effects. In experiments, the SM-SC nanowire is a real spinful electronic 1D Majorana nanowire utilizing a semiconductor with a large Rashba-type SOC in proximity to a conventional $s$-wave superconductor in the presence of a magnetic Zeeman spin splitting field to implement an effective spinless $p$-wave superconductor. The minimal model of a one-dimensional single channel SM-SC nanowire can be described in continuum real space as
\begin{widetext}
    \begin{equation}\label{eq:NW}
            \hat{H}_{\text{NW}}=\int_{0}^{L} \left[\sum_{s,s'}f_{s}^\dagger(x)
             \left(-\frac{\hbar^2\partial_{x}^{2}}{2 m^*}-\mu-i \alpha \sigma_y \partial_x+V_{\text{Z}} \sigma_x\right)_{s,s'} f_{s'}(x)
            +\left( \Delta  f_{\uparrow}^\dagger(x) f_{\downarrow}^\dagger(x) +\text{h.c.} \right) \right]\dd{x},
    \end{equation}
\end{widetext}
where $L$ is the total length of the nanowire, $f_s(x)^\dagger$ [$f_s(x)$] creates (annihilates) an electron at the position $x$ with spin s ($s=\uparrow$/$ \downarrow$), $m^*$ is the effective mass of electrons in the conduction band, $\mu$ is the chemical potential, $\alpha$ is the magnitude of the Rashba-type SOC which is aligned perpendicular to the direction of the 1D nanowire, $V_{\text{Z}}$ is the Zeeman splitting energy arising from the magnetic field applied along the direction of the 1D nanowire, and $\Delta$ is the proximitized $s$-wave SC pairing term. In Hamiltonian~\eqref{eq:NW}, the term in the first parentheses accounts for the semiconductor denoted as $\hat{H}_{\text{SM}}$, and the term in the second parenthesis is for the superconductor denoted as $\hat{H}_{\text{SC}}$. Note that although Eq.~\eqref{eq:NW} is a realistic Hamiltonian for experimental SM-SC hybrid nanowires, it is still a free fermion Hamiltonian within the mean-field BdG description of the induced SC. 

Although the SM-SC nanowire serves as a realistic model of the Kitaev chain to host a pair of Majorana zero modes at the wire ends, the connection between the Kitaev chain and the SM-SC nanowire remains nonobvious. For example, the SM-SC nanowire is established in a spinful system with more terms in the normal (SM) part than the Kitaev chain.
Therefore, we wonder how these two models are connected, e.g., whether there exists any duality between the two models that allows us to map one model into the other. To answer this question, we consider two approaches: 
First, we follow the same procedure as in the Kitaev chain by converting the fermionic Hamiltonian of the SM-SC nanowire into the bosonic spin Hamiltonian through the Jordan-Wigner transformation, and comparing the converted spin model with the transverse-field XY model (which is what the Kitaev chain~\eqref{eq:KC} is formally equivalent to) in the hope of finding some hints for how they might be connected.
Alternatively, we can first project the complete spinful Hamiltonian for the SM-SC nanowire onto an effective low-energy Hamiltonian to obtain an effective spinless system, and then establish the connection since both systems are spinless now.
We will explain the details of the first approach in Sec.~\ref{sec:cluster}, and the second approach in Sec.~\ref{sec:eff}.

Since we establish the Kitaev chain in a lattice model Eq.~\eqref{eq:KC}, we need to study the lattice version of the continuous Hamiltonian for the SM-SC nanowire in Eq.~\eqref{eq:NW} for a direct comparison. The discretization procedure is standard and used extensively. By replacing the differential operator with the finite difference (i.e., $\partial_x f_s^\dagger(x) \mapsto \left[ f_{s}^\dagger\left( x+a\right)-f_{s}^\dagger(x-a) \right]/{2a}$, where $a$ is the fictitious lattice constant in the discretization), we obtain a lattice model for the SM-SC nanowire as
\begin{widetext}
    \begin{equation}\label{eq:NW_2}
        \hat{H}_{\text{NW}}^{(f)}= \sum_{i=1}^{N} \sum_{s,s'} \left[f_{i,s}^\dagger \left( 2t-\mu+V_\text{Z}\sigma_x \right)_{s,s'} f_{i,s'}+ f_{i,s}^\dagger \left( -t-i\alpha_R\sigma_y \right)_{s,s'} f_{i+1,s'} + \text{h.c.}\right] + \left( \Delta f_{i,\uparrow}^\dagger f_{i,\downarrow}^\dagger +\text{h.c.} \right)
    \end{equation}
\end{widetext}
where $f_{i,s}^\dagger=f_{s}^\dagger (ia)$ and $f_{i,s}=f_{s} (ia)$ are the creation and annihilation operators defined at the lattice site $i$, effective nearest neighbor hopping $t=\frac{\hbar^2}{2m^*a^2}$, and reduced SOC $\alpha_R=\frac{\alpha}{2a}$. Here, for a direct comparison between the lattice model for the SM-SC nanowire and the Kitaev chain in Eq.~\eqref{eq:KC}, we impose the periodic boundary condition and drop the factor of $\frac{L}{N}$ before Eq.~\eqref{eq:NW_2} which arises from the discretization.

Like the Kitaev chain, the SM-SC nanowire also carries two topologically distinct phases that cannot be adiabatically connected, which requires us to first solve the band structure of the SM-SC nanowire to find the vanishing of the band gap. Therefore, we rewrite the Hamiltonian~\eqref{eq:NW_2} in momentum space by imposing the periodic boundary condition as 
\begin{equation}\label{eq:NW_k}
    \begin{split}
        &\hat{H}_{\text{NW}}^{(k)}= \sum_{k} \left[ 2t \left( 1- \cos k \right)-\mu \right] \left( \tilde{f}_{k,\uparrow}^\dagger \tilde{f}_{k,\uparrow} + \tilde{f}_{k,\downarrow}^\dagger \tilde{f}_{k,\downarrow} \right) \\
        &+ \left[\left( V_{\text{Z}} - 2i \alpha_R \sin k  \right) \tilde{f}_{k,\uparrow}^\dagger \tilde{f}_{k,\downarrow}  + \Delta \tilde{f}_{k,\uparrow}^\dagger \tilde{f}_{-k,\downarrow}^\dagger  + \text{h.c.}\right]
    \end{split}
    \end{equation}
where $\tilde{f}_{k,s}^\dagger=\frac{1}{\sqrt{N}} \sum_{j} f_{j,s}^\dagger e^{ik j} $, and we set the fictitious lattice constant $a=1$ for notational simplicity. 
To diagonalize Hamiltonian~\eqref{eq:NW_k}, we construct an effective Bogoliubov-de Gennes (BdG) Hamiltonian, $\hat{H}_{\text{NW}}^{(k)}= \sum_{k} \frac{1}{2} \Psi(k)^\dagger H_{\text{NW}}^{\text{BdG}}(k) \Psi(k) + 2t \left( 1- \cos k \right)-\mu$ , where
\begin{equation}
    \begin{split}
        H_{\text{NW}}^{\text{BdG}}(k)&=\left[ 2t\left( 1-\cos k \right) - \mu + V_{\text{Z}}\sigma_x+2\alpha_R \sin k \sigma_y\right] \tau_z\\
        &- \Delta \sigma_y  \tau_y,  
    \end{split}
\end{equation}
and $\Psi(k)=\left( \tilde{f}_{k,\uparrow} ,\tilde{f}_{k,\downarrow} , \tilde{f}_{-k,\uparrow}^\dagger, \tilde{f}_{-k,\downarrow}^\dagger\right)^\intercal$. Here, $\bm{\sigma}$ and $\bm{\tau}$ are vectors of Pauli matrices acting on the spin and particle-hole space respectively. Therefore, the band structure of the SM-SC is 
\begin{equation}\label{eq:NW_Ek}
    \begin{split}
        E_{\text{NW},\pm}^2(k)&=\xi^2(k)+V_{\text{Z}}^2+\Delta^2 + \alpha^2(k) \\
        &\pm 2  \sqrt{V_{\text{Z}}^2\Delta^2+V_{\text{Z}}^2\xi^2(k)+\alpha^2(k)\xi^2(k)},
    \end{split}
\end{equation}
where $\xi(k)=2t\left( 1-\cos k \right)-\mu$, and $\alpha(k)= 2\alpha_R \sin k$. By tracking the vanishing of the band gap at $k=0$, we find that the TQPT happens at $V_{\text{Z}}^2=\mu^2+\Delta^2$ with the trivial (topological) phase being $V_{\text{Z}}^2<\mu^2+\Delta^2$ ($V_{\text{Z}}^2>\mu^2+\Delta^2$) to be contrasted with $2\abs{t}<\abs{\mu}$  (trivial) and $2\abs{t}>\abs{\mu}$ (topological) in the Kitaev chain.

\subsection{Complete spinful model and generalized cluster model}\label{sec:cluster}

We start by directly applying the Jordan-Wigner transformation to the SM-SC nanowire in Eq.~\eqref{eq:NW_2}. First, we need to generalize the previous definition of the Jordan-Wigner transformation for the spinless system to a spinful system with $f_{i,\uparrow} $ and $f_{i,\downarrow} $. 
We define the nonlocal spin operators using the same relation for the Jordan-Wigner transformation Eq.~\eqref{eq:JW} following a designated order as per 
\begin{equation}\label{eq:order}
    f_{1,\uparrow} ~ f_{1,\downarrow} ~ \cdots ~ f_{i,\uparrow} ~ f_{i,\downarrow} ~ f_{i+1,\uparrow} ~ f_{i+1,\downarrow} ~ \cdots ~ f_{N,\uparrow} ~ f_{N,\downarrow}.
\end{equation}
Therefore, the Jordan-Wigner transformation for a spinful fermionic system takes different forms depending on its spin, i.e.,
\begin{equation}\label{eq:JW_spinful}
    \begin{split}
        f_{i,\uparrow}^\dagger &=\prod_{j=1}^{i-1}\left(  \sigma_{j,\uparrow}^{z}\sigma_{j,\downarrow}^{z}  \right)  \sigma_{i,\uparrow}^{+},  \\
        f_{i,\downarrow}^\dagger &=-\prod_{j=1}^{i-1}\left(  \sigma_{j,\uparrow}^{z} \sigma_{j,\downarrow}^{z}  \right)   \sigma_{i,\uparrow}^{z}  \sigma_{i,\downarrow}^{+},  
    \end{split}
\end{equation}
where $ \sigma_{i,\uparrow}^{+} $ and $ \sigma_{i,\downarrow}^{+} $ can be viewed as the raising operators at site $i$ with spin up and down, respectively. 

Since the Hamiltonian for the SM-SC nanowire not only couples the nearest neighbors but also the further sites [``further'' in the sense of the designated order in Eq.~\eqref{eq:order}], e.g., the SOC couples $f_{i,\uparrow}^\dagger $ and $f_{i+1,\downarrow} $, which are separated by a distance of 3, it is worth considering the general form of the spin model arising from the coupling between any two fermionic sites separated by a distance of $\alpha$ in the basis of Majorana operators (due to the existence of the SC term), $\bar{\gamma}_{i,s} \gamma_{i+\alpha,s'}$, where $\gamma_{i,s}= f_{i,s}^\dagger + f_{i,s}   , \bar{\gamma}_{i,s}=i\left( f_{i,s} - f_{i,s}^\dagger \right)$. Following the Jordan-Wigner transformation for spinful fermionic operators in Eq.~\eqref{eq:JW_spinful}, we can define the transformation for Majorana operators as, 
\begin{equation}
    \begin{split}
        \gamma_{i,\uparrow}&=\prod_{j=1}^{i-1}\left( \sigma_{j,\uparrow}^{z}  \sigma_{j,\downarrow}^{z}  \right)  \sigma_{i,\uparrow}^{x},  \\
        \bar{\gamma}_{i,\uparrow}&=\prod_{j=1}^{i-1}\left( \sigma_{j,\uparrow}^{z}  \sigma_{j,\downarrow}^{z}  \right)  \sigma_{i,\uparrow}^{y},\\
        \gamma_{i,\downarrow}&=-\prod_{j=1}^{i-1}\left( \sigma_{j,\uparrow}^{z}  \sigma_{j,\downarrow}^{z}  \right)    \sigma_{i,\uparrow}^{z}  \sigma_{i,\downarrow}^{x},\\
        \bar{\gamma}_{i,\downarrow}&=-\prod_{j=1}^{i-1}\left( \sigma_{j,\uparrow}^{z} \sigma_{j,\downarrow}^{z}  \right)    \sigma_{i,\uparrow}^{z}  \sigma_{i,\downarrow}^{y}.
    \end{split}
\end{equation} 

We summarize in Table~\ref{tab:gamma} that the product of two Majorana operators (in the form of $i\bar{\gamma}\gamma$) that are separated by a distance of $\alpha$ (the index of $\gamma$ minus that of $\bar{\gamma}$) will result in an $(\alpha+1)$-spin interaction in the form of $\sigma^x\underbrace{\sigma^z\dots \sigma^z}_{\abs{\alpha}-1}\sigma^x$ or $\sigma^y\underbrace{\sigma^z\dots \sigma^z}_{\abs{\alpha}-1}\sigma^y$ depending on the sign of $\alpha$, except for $\alpha=0$, which simply corresponds to a polarizing field along $z$ direction.

\begin{table}[ht]
    \begin{ruledtabular}
        \caption{Spin models for the coupling between two Majorana operators separated by a distance of $\alpha$.}
        \begin{tabular}{lcc}
            Fermionic models  & $\alpha$ & Spin models\\
            \hline
            $i \bar{\gamma}_{i+1,\downarrow} \gamma_{i,\uparrow} $ & -3 & $ \sigma_{i,\uparrow}^{y}  \sigma_{i,\downarrow}^{z}  \sigma_{i+1,\uparrow}^{z}  \sigma_{i+1,\downarrow}^{y} $ \\
            $i \bar{\gamma}_{i+1,\uparrow} \gamma_{i,\uparrow} / i \gamma_{i+1,\downarrow} \gamma_{i,\downarrow}  $ & -2 & $  -\sigma_{i,\uparrow}^{y}  \sigma_{i,\downarrow}^{z}  \sigma_{i+1,\uparrow}^{y} / -\sigma_{i,\downarrow}^{y}   \sigma_{i+1,\uparrow}^{z}  \sigma_{i+1,\downarrow}^{y} $ \\
            $i \bar{\gamma}_{i,\downarrow} \gamma_{i,\uparrow} / i \bar{\gamma}_{i+1,\uparrow} \gamma_{i,\downarrow} $ & -1 & $ \sigma_{i,\uparrow}^{y}  \sigma_{i,\downarrow}^{y} / \sigma_{i,\downarrow}^{y}  \sigma_{i+1,\uparrow}^{y} $ \\
            $ i\bar{\gamma}_{i,\uparrow} \gamma_{i,\uparrow} /  i\bar{\gamma}_{i,\downarrow} \gamma_{i,\downarrow}  $  & $0$  &  $\sigma_{i,\uparrow}^{z}/ \sigma_{i,\downarrow}^{z}$  \\
            $ i\bar{\gamma}_{i,\uparrow} \gamma_{i,\downarrow} / i\bar{\gamma}_{i,\downarrow} \gamma_{i+1,\uparrow} $ & $1$ & $ \sigma_{i,\uparrow}^{x}  \sigma_{i,\downarrow}^{x} / \sigma_{i,\downarrow}^{x}  \sigma_{i+1,\uparrow}^{x} $ \\
            $i \bar{\gamma}_{i,\uparrow} \gamma_{i+1,\uparrow} / i \bar{\gamma}_{i,\downarrow} \gamma_{i+1,\downarrow} $ & $2$ & $  -\sigma_{i,\uparrow}^{x}  \sigma_{i,\downarrow}^{z}  \sigma_{i+1,\uparrow}^{x} / -\sigma_{i,\downarrow}^{x}   \sigma_{i+1,\uparrow}^{z}  \sigma_{i+1,\downarrow}^{x} $ \\
            $i \bar{\gamma}_{i,\uparrow} \gamma_{i+1,\downarrow} $ & 3 & $  \sigma_{i,\uparrow}^{x}   \sigma_{i,\downarrow}^{z}  \sigma_{i+1,\uparrow}^{z}  \sigma_{i+1,\downarrow}^{x} $ 
        \end{tabular}
    \label{tab:gamma}
    \end{ruledtabular}
\end{table}

Therefore, owing to the second term in Eq.~\eqref{eq:NW_2} describing the hopping between $f_{i,s}^\dagger $ and $f_{i+1,s'}^\dagger $, we expect many spin-interactions, i.e., $n$-spin interactions with $n\ge 2$, which looks superficially similar to the cluster model~\cite{suzuki1971relationship,briegel2001persistent}. 
Therefore, to obtain the spin model corresponding to Eq.~\eqref{eq:NW_2}, we first rewrite Eq.~\eqref{eq:NW_2} in terms of Majorana operators by replacing $f_{i,s}^\dagger  f_{i,s}$ with $\left( 1+i \bar{\gamma}_{i,s} \gamma_{i,s} \right)/2$,  $f_{i,s}^\dagger  f_{j,s'}+\text{h.c.}$ with $i \left( \bar{\gamma}_{i,s} \gamma_{j,s'} + \bar{\gamma}_{j,s'} \gamma_{i,s}  \right)/2$ [$(i,s)\neq (j,s')$], and $f_{i,s}^\dagger f_{j,s'}^\dagger + \text{h.c.} $ with $i\left( \bar{\gamma}_{i,s} \gamma_{j,s'} - \bar{\gamma}_{j,s'} \gamma_{i,s}  \right)$ [$(i,s)\neq (j,s')$]. The Hamiltonian in the Majorana basis is still simply quadratic (because Hamiltonian~\eqref{eq:NW_2} is indeed a free fermion problem in the mean-field level) as per
\begin{equation}\label{eq:NW_gamma}
    \begin{split}
        &\hat{H}_{\text{NW}}^{(\gamma)}=\left( 2t-\mu \right)N+\frac{i}{2}\sum_{i=1}^{N}\\
        &{\left( 2t-\mu \right)}\left( \bar{\gamma}_{i,\uparrow} \gamma_{i,\uparrow} + \bar{\gamma}_{i,\downarrow} \gamma_{i,\downarrow}   \right) +  V_{\text{Z}} \left(\bar{\gamma}_{i,\uparrow} \gamma_{i,\downarrow} + \bar{\gamma}_{i,\downarrow} \gamma_{i,\uparrow}  \right)\\
        &-t \left(  \bar{\gamma}_{i,\uparrow} \gamma_{i+1,\uparrow}  +  \bar{\gamma}_{i,\downarrow} \gamma_{i+1,\downarrow} +  \bar{\gamma}_{i+1,\uparrow} \gamma_{i,\uparrow} +  \bar{\gamma}_{i+1,\downarrow} \gamma_{i,\downarrow} \right)\\
        &-\alpha_R\left(  \bar{\gamma}_{i,\uparrow} \gamma_{i+1,\downarrow} +  \bar{\gamma}_{i+1,\downarrow} \gamma_{i,\uparrow} - \bar{\gamma}_{i,\downarrow} \gamma_{i+1,\uparrow} -   \bar{\gamma}_{i+1,\uparrow} \gamma_{i,\downarrow}  \right)\\
        &+\Delta\left(  \bar{\gamma}_{i,\uparrow} \gamma_{i,\downarrow} - \bar{\gamma}_{i,\downarrow} \gamma_{i,\uparrow}  \right) .
    \end{split}
\end{equation} 

Next, we refer to Table~\ref{tab:gamma}, and substitute all products of Majorana operators in Eq.~\eqref{eq:NW_gamma} to obtain the Hamiltonian for the corresponding bosonic spin model for the SM-SC nanowire as per
\begin{equation}\label{eq:NW_JW}
    \begin{split}
        &\hat{H}_{\text{NW}}^{(\sigma)}=\left( 2t-\mu \right)N+\frac{1}{2}\sum_{i=1}^{N}\\
        & \left( 2t-\mu \right)\left(  \sigma_{i,\uparrow}^{z} + \sigma_{i,\downarrow}^{z}  \right)+\left( V_{\text{Z}}+\Delta \right) \sigma_{i,\uparrow}^{x}  \sigma_{i,\downarrow}^{x} +\left( V_{\text{Z}}-\Delta \right) \sigma_{i,\uparrow}^{y}  \sigma_{i,\downarrow}^{y} \\
        &+ \alpha_R \left(  \sigma_{i,\downarrow}^{x}  \sigma_{i+1,\uparrow}^{x} + \sigma_{i,\downarrow}^{y}  \sigma_{i+1,\uparrow}^{y}  \right) + t \left( \sigma_{i,\downarrow}^{x}  \sigma_{i+1,\uparrow}^{z}  \sigma_{i+1,\downarrow}^{x} \right.\\
        & \left. + \sigma_{i,\downarrow}^{y}  \sigma_{i+1,\uparrow}^{z}  \sigma_{i+1,\downarrow}^{y} + \sigma_{i,\uparrow}^{x}  \sigma_{i,\downarrow}^{z}  \sigma_{i+1,\uparrow}^{x} + \sigma_{i,\uparrow}^{y}  \sigma_{i,\downarrow}^{z}  \sigma_{i+1,\uparrow}^{y} \right)\\
        &- \alpha_R\left(  \sigma_{i,\uparrow}^{x}  \sigma_{i,\downarrow}^{z}  \sigma_{i+1,\uparrow}^{z}  \sigma_{i+1,\downarrow}^{x} +  \sigma_{i,\uparrow}^{y}  \sigma_{i,\downarrow}^{z}  \sigma_{i+1,\uparrow}^{z}  \sigma_{i+1,\downarrow}^{y}  \right) .
    \end{split}
\end{equation}

Equation~\eqref{eq:NW_JW} is essentially a generalized cluster model with staggered couplings if we relabel the indices of spin operators according to the order in Eq.~\eqref{eq:order}.  By redefining $ \sigma_{i,\uparrow}^{x,y,z} = \sigma_{2i-1}^{x,y,z}$ and $ \sigma_{i,\downarrow}^{x,y,z} = \sigma_{2i}^{x,y,z}$, the generalized cluster model becomes obvious as 
\begin{widetext}
    \begin{equation}\label{eq:NW_cluster}
        \hat{H}_{\text{C}}^{(\sigma)}=\left( 2t-\mu \right)N+\sum_{i=1}^{2N} \left( J_{1,i} \sigma_{i}^{z} + \sum_{s=\{x,y\}}J_{2,i}^{s} \sigma_{i}^{s} \sigma_{i+1}^{s}  + J_{3,i}^s \sigma_{i}^{s} \sigma_{i+1}^{z} \sigma_{i+2}^{s}+ J_{4,i}^s \sigma_{i}^{s} \sigma_{i+1}^{z} \sigma_{i+2}^{z} \sigma_{i+3}^{s} \right)  ,
    \end{equation} 
with constant couplings for $J_{1,i}^{x,y}$ and $J_{3,i}^{x,y}$, and the staggered couplings for $J_{2,i}^{x,y}$ and $J_{4,i}^{x,y}$ 
    \begin{equation}\label{eq:NW_J}
        J_{1,i}=\frac{2t-\mu}{2},~ J_{2,i}^{x}=\begin{cases}
            \frac{\alpha_R}{2} & i\in2\mathbb{Z}^+\\
            \frac{V_{\text{Z}}+\Delta}{2} & i\in2\mathbb{Z}^+-1
    \end{cases},
    ~  J_{2,i}^{y}=\begin{cases}
        \frac{\alpha_R}{2} & i\in2\mathbb{Z}^+\\
        \frac{V_{\text{Z}}-\Delta}{2} & i\in2\mathbb{Z}^+-1
    \end{cases},
    ~ J_{3,i}^{x}=J_{3,i}^{y}=\frac{t}{2},
    ~ J_{4,i}^{x}=J_{4,i}^{y}=\begin{cases}
        0 & i\in 2\mathbb{Z}^+\\
        -\frac{\alpha_R}{2} & i\in 2\mathbb{Z}^+-1
    \end{cases}.
\end{equation}
\end{widetext}

Due to the 3-spin and 4-spin interactions, we see that the spin model for the SM-SC through a direct Jordan-Wigner transformation is much more complicated than that of the Kitaev chain, i.e., a transverse-field XY model in Eq.~\eqref{eq:XY} which only contains the polarizing term and 2-spin interactions. Thus, one key formal difference between the Kitaev chain and Majorana nanowire is that while the former becomes more transparent theoretically through the Jordan-Wigner transformation by mapping into a spin problem well-understood for more than 50 years~\cite{pfeuty1970onedimensional}, the latter is more transparent, in fact, in the original fermionic picture, as its spin dual model is essentially an unknown and complex spin cluster model.

For a generalized cluster model in $1$D, it has been shown~\cite{verresen2017onedimensional} that a chain composed of even-number spin interactions (in the form of $\sigma^x\sigma^z\dots \sigma^z\sigma^x$ ) has $\mathbb{Z}_2$ symmetry breaking (SB) while a chain composed of odd-number spin interactions can carry symmetry-protected topological (SPT) phases. However, in our spin model in Eqs.~\eqref{eq:NW_cluster} and~\eqref{eq:NW_J}, i.e., with a collection of both even number spin interactions and odd number spin interactions with staggered spin couplings, the connection to the SPT phases or SB phase, and how the transition happens are not obvious in the spin language.

\subsection{Effective low-energy spinless model and Kitaev chain}\label{sec:eff}

Although the direct application of the Jordan-Wigner transformation to the SM-SC nanowire does not provide a mapping to the Kitaev chain due to the 3-spin and 4-spin interactions, we can see their connection by projecting the spinful SM-SC nanowire onto an effective spinless low-energy model. 
The logic is that: (1) we first solve the band structure of the SM part in Hamiltonian~\eqref{eq:NW_2} by rewriting it in momentum space; (2) then we obtain an effective low-energy model by projecting out the upper band with higher energy; (3) finally, we rewrite the projected effective low-energy Hamiltonian back into the real space and compare that with the Kitaev chain. The logic is based on the idea that eliminating the spinfulness of the Majorana nanowire in a systematic manner converting it into an effective low-energy spinless system should provide a direct clue to its connection with the spinless Kitaev chain.

To solve the SM (normal) term $\hat{H}_{\text{SM}}^{(k)}$ (obtained by setting $\Delta=0$ in Eq.~\eqref{eq:NW_k}) as the first step, we can directly diagonalize it to obtain the band structure of the bare SM, $E_{\text{SM},\pm}(k)=2t\left( 1-\cos k \right)-\mu\pm \sqrt{4\alpha_R^2\sin^2 k+V_{\text{Z}}^2}$. Therefore, the Hamiltonian for the SM part can be reconstructed as
\begin{equation}\label{eq:HSM}
    \hat{H}_{\text{SM}}^{(k)}=\sum_{k} E_{+}(k) \tilde{f}_{k,+}^\dagger \tilde{f}_{k,+} + E_{-}(k) \tilde{f}_{k,-}^\dagger \tilde{f}_{k,-},
\end{equation}
where $\tilde{f}_{k,+}^\dagger$ and $\tilde{f}_{k,-}^\dagger$ create an electron in the upper and lower bands, respectively, and are connected to the original (physical) $\tilde{f}_{k,\uparrow}^\dagger$ and $\tilde{f}_{k,\downarrow}^\dagger$ through the unitary transformation
\begin{equation}\label{eq:fk}
    \begin{pmatrix}
        \tilde{f}_{k,+}^\dagger \\
        \tilde{f}_{k,-}^\dagger 
    \end{pmatrix}=
    \frac{1}{\sqrt{2}}
    \begin{pmatrix}
        1 & e^{i\phi(k)}\\
        e^{-i\phi(k)} & -1 
    \end{pmatrix}
    \begin{pmatrix}
        \tilde{f}_{k,\uparrow}^\dagger \\
        \tilde{f}_{k,\downarrow}^\dagger 
    \end{pmatrix},
\end{equation}
where $\phi(k)=\arctan(\dfrac{2\alpha_R\sin k}{V_{\text{Z}}})$.

Since we are only interested in low-energy physics, we can project out the upper band $E_{+}(k)$ by simply setting $\tilde{f}_{k,+}^\dagger$ to 0. Thus, the effective low-energy model can be obtained by substituting Eq.~\eqref{eq:fk} into the SC part of Hamiltonian~\eqref{eq:NW_k} as
\begin{equation}\label{eq:NW_eff}
    \begin{split}
        \hat{H}_{\text{eff}}^{(k)} &= \sum_k \left[ 2t\left( 1-\cos k \right)-\mu- \sqrt{4\alpha_R^2\sin^2 k+V_{\text{Z}}^2} \right] \tilde{f}_{k}^\dagger \tilde{f}_{k} \\
        &+ \frac{2i\Delta \alpha_R \sin k}{\sqrt{(2\alpha_R\sin k)^2+V_{\text{Z}}^2}} \left( \tilde{f}_{k}^\dagger \tilde{f}_{-k}^\dagger + \tilde{f}_{k} \tilde{f}_{-k}  \right),
    \end{split}
\end{equation}
where we drop the band index $\pm$ in fermionic operators from now on due to the elimination of the upper band. Thus, the effective low-energy model Eq.~\eqref{eq:NW_eff} can be also viewed as a spinless system, which takes a more similar form to the Kitaev chain.

However, such a projection has implicitly assumed the large $V_{\text{Z}}$ limit in order to minimize the occupancy on the upper band $E_{\text{SM},+}(k)$, which further indicates the deep topological regime. For smaller $V_\text{Z}$, the effective low-energy model fails to faithfully characterize the band bottom of $E_{\text{NW},-}(k)$ near zero momentum, which is where the band gap vanishes. Therefore, this low-energy model is not able to reproduce the real TQPT at $V_{\text{Z}}^2=\mu^2+\Delta^2$. 
Despite this low-energy model being unsuitable to describe the physics of the SM-SC nanowire in all regimes, we may still treat the effective low-energy model as an independent model from a purely theoretical viewpoint by ignoring the fact that this approximation relies on large $V_{\text{Z}}$. 

Therefore, we obtain the band structure and track the vanishing of the band gap by writing down the Bogoliubov-de Gennes Hamiltonian for Eq.~\eqref{eq:NW_eff},  $\hat{H}_{\text{eff}}^{(k)}= \frac{1}{2}\sum_{k}\tilde{\bm{f}}_{k}^\dagger  H_{\text{eff}}^{\text{BdG}}(k) \tilde{\bm{f}}_{k} + 2t\left( 1-\cos k \right)-\mu-\sqrt{4\alpha_R^2\sin^2k+V_{\text{Z}}^2}$, where $\tilde{\bm{f}}_{k} =\left( \tilde{f}_{k} , \tilde{f}_{-k}^\dagger  \right)^\intercal$, and $H_{\text{eff}}^{\text{BdG}}(k)=\bm{h}(k)\cdot \bm{\tau}$. Here, $\bm{h}(k)=\left( h_x(k),h_y(k),h_z(k) \right) $ where 
\begin{equation}
    \begin{split}
        h_x(k)&=0,\\
        h_y(k)&=-\frac{4\Delta \alpha_R \sin k}{\sqrt{(2\alpha_R\sin k)^2+V_{\text{Z}}^2}},\\
        h_z(k)&= 2t\left( 1-\cos k \right)-\mu- \sqrt{4\alpha_R^2\sin^2 k+V_{\text{Z}}^2} .
    \end{split}    
\end{equation} 
Therefore, the band structure is simply $E_{\text{eff}, \pm}(k)=\pm \sqrt{h_y(k)^2+h_z^2(k)}$, and band gap vanishes when $V_{\text{Z}}=-\mu$ at $k=0$ or $V_{\text{Z}}=4t-\mu$ at $k=\pi$, which divides the phase diagram into three topologically distinctive regimes ($V_{\text{Z}}<-\mu$, $-\mu<V_{\text{Z}}<4t-\mu$, and $V_{\text{Z}}>4t-\mu$).  

To study the topological property in each regime, we define the winding number $ \mathcal{W}=\frac{1}{2\pi}\oint d \theta(k)$ in the $h_y$-$ h_z$ plane as the topological invariant, where $\theta(k)=\arctan(h_z(k)/h_y(k))$, which is further equivalent to a simpler form 
\begin{equation}\label{eq:W}
    \mathcal{W}=\text{sign}\left( h_z(0) h_z(\pi) \right).
\end{equation}
Therefore, by applying Eq.~\eqref{eq:W} to the three aforementioned regimes, we find two phases: the topological phase with $\abs{\mathcal{W}}=1$ for $-\mu<V_{\text{Z}}<4t-\mu$, and the trivial phase with $\abs{\mathcal{W}}=0$ for $V_{\text{Z}}>4t-\mu$ or $V_{\text{Z}}<-\mu$. We note that the trivial regime for $V_{\text{Z}}>4t-\mu$ can be eliminated in practice by setting a sufficiently small fictitious lattice constant $a$. This second trivial phase appears to be an artifact of discretization. So our statement of large $V_{\text{Z}}$ indicating deep topological regime still holds in the ideal case, where $t$ approaches infinity. Therefore, the ``physical'' phase boundary is just $V_{\text{Z}}=-\mu$, which is different from the original SM-SC nanowire ($V_{\text{Z}}^2=\mu^2+\Delta^2$) because of the inaccurate projection from the complete spinful Hamiltonian~\eqref{eq:NW_k} onto the effective low-energy spinless Hamiltonian~\eqref{eq:NW_eff} near the real TQPT. Thus, we will focus on the model deep inside the topological regime with a reasonably good approximation of the projection, and restore the real space Hamiltonian to uncover the connection between the SM-SC nanowire and Kitaev chain by establishing a direct mapping.

So far, from the effective low-energy model in momentum space in Eq.~\eqref{eq:NW_eff}, we can already tell that it is a spinless $p$-wave superconductor with a normal part in the first line, and an antisymmetric SC part in the second line. However, there are still some slight differences between Eq.~\eqref{eq:NW_eff} and the Kitaev chain in Eq.~\eqref{eq:KC_k}, e.g., Equation~\eqref{eq:NW_eff} contains a much more complicated form of the $p$-wave superconducting pairing than Eq.~\eqref{eq:KC_k}.
To see the underlying connection between the two models, we need to obtain the effective low-energy spinless model in real space by expanding the Hamiltonian~\eqref{eq:NW_eff} (i.e., $\sqrt{4\alpha_R^2\sin^2 k+V_{\text{Z}}^2}$ in the normal part and $\frac{2i\Delta \alpha_R \sin k}{\sqrt{(2\alpha_R\sin k)^2+V_{\text{Z}}^2}}$ in the SC part) near  $\frac{\alpha_R}{V_{\text{Z}}}\sim0$ (assuming the large $V_{\text{Z}}$ limit) to the $n$-th order as shown the second column in Table~\ref{tab:H}. From the form of the expansion, we notice the long-range hoppings and pairings arising from high-frequency modes (e.g., $\cos 2k, \sin 3k$) for the expansion with an order $>$ 2. By performing the inverse Fourier transformation, we restore the effective low-energy spinless model in real space as shown in the third column in Table~\ref{tab:H}. 

In Table~\ref{tab:H}, the zeroth order approximation is simply a trivial SM without the SC pairing term, which is not of any interest. However, more interesting mapping emerges once we go to the first order, where the Hamiltonian $\hat{H}_{\text{eff},1}^{(r)}$ becomes an exact Kitaev chain if we map $(\mu_{\text{KC}},t_{\text{KC}},\Delta_{\text{KC}})\mapsto(\mu-2t+V_{\text{Z}},t,-\frac{2\Delta\alpha_R}{V_{\text{Z}}})$.
The TQPT of this mapped Kitaev chain from the first order approximation is $\abs{\mu-2t+V_{\text{Z}}}=2t$ (as $t=\frac{\hbar^2}{2m^*a^2}>0$ ), which also coincides with the TQPT of the original effective low-energy spinless model~\eqref{eq:NW_eff}. 
Therefore, the Kitaev chain can be at least viewed as a first-order approximation of the SM-SC nanowire deep inside the topological regime.
For the higher orders of approximation, more detailed corrections to the normal (SC) part are added at an even (odd) order of approximation, which corresponds to the Kitaev chain with a longer range of hoppings and pairings. In principle, the original spinless effective low-energy model Eq.~\eqref{eq:NW_eff} should correspond to a real space model with infinite couplings. 
However, as opposed to the general Kitaev chain with long-range couplings, which can carry multiple pairs of MZMs~\cite{degottardi2013majoranaa,alecce2017extended}, all orders of Kitaev chains in Table~\ref{tab:H} are topologically equivalent because their resulting band structures are all adiabatically connected due to the exact vanishing of the high-frequency mode at $k=0$ or $k=\pi$ (which is also where the band inversion happens). 
{In principle, one can have multiple pairs of MZMs in Majorana nanowires if one allows the multi-subband occupancy~\cite{stanescu2011majorana,lutchyn2011search} (which is actually a more experimentally realistic model), because Hamiltonian~\eqref{eq:NW} can be constructed completely real in 1D (as long as Zeeman field is parallel to the nanowire as it is now), leading to an effective chiral symmetry in class BDI characterized by $\mathbb{Z}$ invariant~\cite{schnyder2008classification,kitaev2009periodic,ryu2010topological} in 1D system by redefining the time-reversal operator to a mere complex conjugate $\mathcal{K}$~\cite{tewari2012topological}. 
Therefore, the multi-subband Hamiltonian Eq.~\eqref{eq:NW} with $N$ pairs of MZMs is equivalent to $N$ occupied Kitaev chains Eq.~\eqref{eq:KC} along with $N$ unoccupied Kitaev chains being energetically pushed upward by the sufficiently large Zeeman splitting field (these $N$ unoccupied bands are also the bands that we projected out as shown in Eqs.~\eqref{eq:HSM} and~\eqref{eq:fk}), which can be further considered as a single Kitaev chain with long-range hoppings and pairings in class BDI in 1D due to the real SC pairings in the Hamiltonian~\eqref{eq:KC}.
However, in this paper, because we restrict the Majorana nanowire Eq.~\eqref{eq:NW} in the strict one subband limit, we can only have one pair of MZMs on both ends in the Majorana nanowire, topologically equivalent to the case of a single Kitaev chain where next-nearest-neighbor couplings along with all further couplings in the Kitaev chain vanish~\cite{niu2012majorana}, which can only carry at most one pair of MZMs.}
Furthermore, the fact that the leading terms in the long-range couplings in Table~\ref{tab:H} decay exponentially as $\left( \frac{\alpha_R}{V_{\text{Z}}}\right)^d$ with $d$ being the distance between two coupled sites indicates that this model should essentially be thought of as a short-range model equivalent to the Kitaev chain with only the nearest-neighbor hoppings and pairings. Therefore, all effective low-energy spinless models in Table~\ref{tab:H} have the same TQPT, and thus the same phase diagram. Their essential physics of a spinless $p$-wave topological superconductor is unchanged despite the formal differences in terms of the range of hoppings and pairings in the Hamiltonian at different orders of approximation.

\begin{widetext}

\begin{table}[ht]
    \begin{ruledtabular}
        \caption{Effective low-energy spinless models up to $n$-th order approximation.}
        \begin{tabular}{lcc}
            $n$ & $k$-space $\hat{H}_{\text{eff},n}^{(k)} = \sum_{k}\dots $ & $r$-space $\hat{H}_{\text{eff},n}^{(r)}=\sum_{i=1}^L \dots $ \\
            \hline
            0 & $\left[ 2t\left( 1-\cos k \right)-\mu-V_{\text{Z}} \right] \tilde{f}_{k}^\dagger \tilde{f}_{k} $  & $ \left( 2t-\mu-V_{\text{Z}} \right) f_{i}^\dagger  f_{i} -t \left( f_{i}^\dagger  f_{i+1} +\text{h.c.} \right)$ \\
            1 & $\left[ 2t\left( 1-\cos k \right)-\mu-V_{\text{Z}} \right] \tilde{f}_{k}^\dagger \tilde{f}_{k} + \left(\frac{2i\Delta\alpha_R\sin k}{V_{\text{Z}}} \tilde{f}_{k}^\dagger \tilde{f}_{-k}^\dagger +\text{h.c.} \right)$ &  $ \left( 2t-\mu-V_{\text{Z}} \right) f_{i}^\dagger  f_{i} + \left( -t f_{i}^\dagger  f_{i+1} + \frac{2\Delta\alpha_R}{V_{\text{Z}}}f_{i}^\dagger f_{i+1}^\dagger +\text{h.c.}\right)$ \\
            2 & $\mqty{\left[ 2t\left( 1-\cos k \right)-\mu- V_{\text{Z}} - \frac{2\alpha_R^2\sin^2k}{V_{\text{Z}}}\right] \tilde{f}_{k}^\dagger \tilde{f}_{k} \\
            + \left(\frac{2i\alpha_R\Delta\sin k}{V_{\text{Z}}}  \tilde{f}_{k}^\dagger \tilde{f}_{-k}^\dagger + \text{h.c.}  \right)}$ & $\mqty{ \left( 2t-\mu-V_{\text{Z}} - \frac{\alpha_R^2}{V_{\text{Z}}}\right) f_{i}^\dagger  f_{i} + \left( -t f_{i}^\dagger  f_{i+1}+ \frac{\alpha_R^2}{2V_{\text{Z}}} f_{i}^\dagger  f_{i+2}  +\text{h.c.} \right)\\
            +\left( \frac{2\Delta\alpha_R}{V_{\text{Z}}} f_{i}^\dagger f_{i+1}^\dagger+\text{h.c.} \right)}$ \\
            3 & $\mqty{\left[ 2t\left( 1-\cos k \right)-\mu- V_{\text{Z}} - \frac{2\alpha_R^2\sin^2k}{V_{\text{Z}}}\right] \tilde{f}_{k}^\dagger \tilde{f}_{k} \\+ \left[\left( \frac{2i\alpha_R\Delta\sin k}{V_{\text{Z}}} - \frac{4i\alpha_R^3\Delta\sin^3k}{V_{\text{Z}}^3}  \right)  \tilde{f}_{k}^\dagger \tilde{f}_{-k}^\dagger + \text{h.c.}  \right]}$ & $\mqty{ \left( 2t-\mu-V_{\text{Z}} - \frac{\alpha_R^2}{V_{\text{Z}}}\right) f_{i}^\dagger  f_{i}  +\left( -tf_{i}^\dagger  f_{i+1}+ \frac{\alpha_R^2}{2V_{\text{Z}}} f_{i}^\dagger  f_{i+2} +\text{h.c.} \right) \\
            +\left[\left( \frac{2\alpha_R\Delta}{V_{\text{Z}}}- \frac{3\alpha_R^3\Delta}{V_{\text{Z}}^3}\right) f_{i}^\dagger f_{i+1}^\dagger + \frac{\Delta\alpha_R^3}{V_{\text{Z}}^3} f_{i}^\dagger f_{i+3}^\dagger +\text{h.c.} \right] }$ \\
        \end{tabular}
    \label{tab:H}
    \end{ruledtabular}
\end{table}
\end{widetext}

\section{Conclusion}\label{sec:conclusion}

In this paper, we study the underlying connection between the Kitaev chain toy model and the SM-SC realistic Majorana nanowire model, motivated by the fact that both of these 1D systems can host Majorana zero modes at the boundaries in the open geometry. We first establish the known duality between the Kitaev chain and the transverse-field XY model (which can be further fine-tuned to a transverse-field Ising model under a specific set of parameters) through the Jordan-Wigner transformation, and then ask the question what the corresponding spin dual model for the SM-SC nanowire looks like if we apply the Jordan-Wigner transformation directly. Following this logic, we eventually find that, unlike the Kitaev chain, which can be transformed into a well-known spin model with only 2-spin interactions, the SM-SC nanowire corresponds to a complicated dual generalized spin cluster model with staggered couplings, which indicates that the connection between the Kitaev chain and SM-SC nanowire is not \textit{a priori} obvious.

Therefore, we resort to a different approach, which is to project out the upper band of the bare SM with higher energy to obtain an effective low-energy spinless nanowire model in the large $V_{\text{Z}}$ limit. This spinless model in momentum space is already a spinless $p$-wave SC, which can be further Fourier transformed into the real space to show a direct mapping to the original Kitaev chain. Finally, we find that the Kitaev chain can be at least viewed as a first-order approximation of the Majorana nanowire deep in the topological regime. The two models are identical deep in the MZM-carrying topological regime. However, the real TQPT cannot be faithfully reproduced in this effective low-energy model as the projection from a complete spinful model to the effective low-energy spinless model is only accurate in the large $V_{\text{Z}}$ limit.

{We emphasize that the statement that the Kitaev chain is isomorphic to the nanowire model~\cite{lutchyn2010majorana}, as is often claimed in the literature, is incorrect or at least incomplete since the Kitaev chain is spinless and the nanowire is spinful, and one cannot thus obtain the nanowire model ever from the Kitaev chain. However, the
opposite statement that the Kitaev chain may be a low-energy projection of the nanowire model into a single spin subband could be correct in some situations. In the this paper, we explicitly
point out when this is true, and when it is not, and the necessary conditions for this restricted isomorphism.  In general, the two models are very different except that they belong to the same Ising anyon universality class for the topological quantum critical transition between the trivial and the topological phases.}

In principle, we can find more quantitatively accurate projections considering the SC term as well by constructing a pseudo ``time-reversal'' symmetry as a function of momentum to block-diagonalize the complete spinful SM-SC BdG Hamiltonian such that spin-up and spin-down channels do not couple. However, an extra degree of freedom then emerges as one can separate the normal part and SC part arbitrarily, leading to non-unique effective models. But once we select one effective model among them, we again have a spinless model which can be used to compare with the Kitaev chain throughout the trivial and topological regime (not just deep in the topological regime). However, the choice of this gauge as a function of $k$ is a nontrivial question, which is left for future work. In fact, it is unclear to us on physical grounds (i.e., the assumption of spinless $p$-wave SC in the Kitaev chain) that it is possible to find a complete theoretical equivalence between these two models except deep in the topological regime.

\section*{Acknowledgment}
We thank Jay Deep Sau, Fengcheng Wu, Jiabin Yu, Yu-An Chen, Fnu Setiawan, Chun-Xiao Liu, and Dinhduy Vu for helpful discussions. This paper is supported by the Laboratory for Physical Sciences. 

\bibliography{Paper_JW}
\appendix

\end{document}